# TaSer (TabAnno and SeqMiner): a toolset for annotating and querying next-generation sequence data

Xiaowei Zhan[1+,*] Dajiang J. Liu[1+,*]

[1]Department of Biostatistics, Center of Statistical Genetics, University of Michigan, Ann Arbor, MI, 48109

**ABSTRACT**

**Summary:** We develop *TaSer* (*TabAnno* and *SeqMiner*), a toolkit for annotating and querying next generation sequence (NGS) dataset in tab-delimited files. *TabAnno* is a powerful and efficient command-line tool designed to pre-process sequence data, annotate variations and generate an indexed feature-enriched project file that can integrate multiple sources of information. Using the project file generated by *TabAnno*, complex queries to the sequence dataset can be performed using *SeqMiner*, an R-package designed to efficiently access large datasets. Extracted information can be conveniently viewed and analyzed by tools in R. *TaSer* is optimized and computationally more efficient than software using database systems. It enables annotating and querying NGS dataset using moderate computing resource.

**Availability and implementation:** *TabAnno* can be downloaded from github (zhanxw.github.io/anno/). *SeqMiner* is distributed on CRAN (cran.r-project.org/web/packages/seqminer).

**Contact**: X.Z. (zhanxw@umich.edu) D.J.L (dajiang@umich.edu)

## 1   INTRODUCTION

Large amount of next generation sequence (NGS) data have been generated using next generation sequencing, in order to get a more detailed understanding of human genomic variations. The analysis of NGS data poses formidable computational challenges. Datasets of these sequence variations [e.g. in variant call format (VCF) (Danecek, et al., 2011)] can be very large in size, even after compression. It is usually impossible to load an entire file into computer memory. Querying information for a genomic region of interest can be challenging.

One standard approach to manipulate genetic data file is to use database management system (DBMS) (San Lucas, et al., 2012), where a database project will need to be built for a NGS dataset to facilitate complex queries. Although DBMS is very powerful for annotating and querying variants, building and updating projects for large NGS datasets require considerable computational resources. Even with a high-end computer server, it may still be necessary to divide large files, and build multiple smaller projects, in order to complete the analysis. For simple interval queries, an alternative tool *tabix* (Li, 2011) can be used. *Tabix* is computationally efficient, but cannot be directly applied to make more complex queries, e.g. getting genotype information for functionally damaging variants in a gene/pathway.

To overcome limitations of existing tools, we developed *TaSer* (*TabAnno* and *SeqMiner*), a new toolset for annotating and querying sequence datasets, which combines the flexibility of DBMS and the efficiency of *tabix*. Our workflow starts by pre-processing the sequence dataset using *TabAnno*, a command line annotator, and generating a feature-enriched index project file that integrates annotation information of genetic variants and optionally external bioinformatics databases, e.g. pre-calculated PolyPhen scores (Adzhubei, et al., 2010). Using the indexed project file generated by *TabAnno*, complex queries can be performed via *SeqMiner*, an R-package that integrates the *tabix* library and supports efficient random access to large datasets. Extracted information is available as standard R objects. Subsequent analysis of the dataset can be conveniently performed in R.

Preprocessing and annotating NGS dataset using *TaSer* is much faster and memory efficient than building database projects. Additionally, our tools are integrated with R, an interactive/programmable environment, so it can be more powerful and flexible than DBMS for performing downstream statistical analysis, visualization, etc.

We compared the performance of *TaSer* with *varianttools*, a DBMS-based tool for annotating and extracting variants. We showed that *TaSer* are more time and memory efficient and can be applied to handle large NGS datasets with moderate computing resources. We also compared *TaSer* with *VariantAnnotation*, an R-package with similar functions. We demonstrate that *TaSer* is much faster, simpler to use and offers unique functionalities, such as calculating summary metrics (e.g. transition/transversion ratio etc.) for the entire NGS dataset. Additionally, *TaSer* also supports more input file formats than both competing tools, which saves user's effort for preparing intermediate files.

## 2   DESCRIPTION

### 2.1   Annotate Sequence Datasets Using *TabAnno*

*TabAnno* is a tool for annotating sequence variants and integrating multiple sources of data. Given that standard annotation only requires chromosomal positions, reference and alternative alleles, *TabAnno* is designed to support generic tab-delimited files as input [e.g. VCF files of genotype calls or METAL (Willer, et al., 2010) files of summary statistics], which saves user efforts for preparing intermediate files.

*TabAnno* supports standard gene-based annotation via commonly used gene/transcript definition, including refFlat or UCSC KnownGenes. Specifically, to annotate each mutation, a reading frame is first determined by the transcripts where the mutation lies. We then obtain the codons before and after the mutation from the reference genome. Synonymous/non-synonymous variants will be annotated by whether or not the mutation induces changes on the amino acid, according to the universal genetic code. It can also annotate genomic regions of interest, e.g. regions overlapping transcription factor binding sites. Specifically, external information defin-ing the region of interest can be stored as BED files (Kent, et al., 2002), and used in *TabAnno* for annotations. A variety of bioinformatics databases are supported and can be incorporated in the annotation, e.g. pre-computed PolyPhen or GERP scores (Cooper, et al., 2005). More Detailed features and usage of *TabAnno* can be found on the authors' website.

---
[*]To whom correspondence should be addressed.
[+]These authors contributed equally

Several optimizations were implemented in *TabAnno* to enable efficient access to large datasets: 1.) It can read/write bgzip compressed files, which minimizes disk I/O, a major bottleneck for processing high throughput data. 2.) We pre-processed external databases via compression and indexing. For example, we store genomic regions (e.g. transcription factor binding sites) as ordered array. Consequently, each record can be retrieved with a time complexity of $O(\log(n))$, where n is the number of regions.

## 2.2 Perform Complex Queries with *SeqMiner*

Using the project file created by *TabAnno*, complex queries to NGS datasets can be performed via *SeqMiner*, an R-package that integrates *tabix*, naturally inherits all its benefits and allows random access to feature-enriched tab-delimited files. Retrieved information is stored in standard R data objects, (e.g. matrix or list). Subsequent data quality control (QC), visualization, analysis can be conveniently performed in R.

One major function for *SeqMiner* is to query compressed and indexed VCF files. With built-in functions, user can conveniently extract genotype information for variants that reside in a given gene, or belong to a certain mutation type (e.g. non-synonymous). Users can also specify and extract additional fields, such as genotype likelihoods, or read depths, etc.

In addition to VCF files, *SeqMiner* can support queries to generic tab-delimited files. For example, our tools can query METAL files and extract association test statistics for variants in a given gene, etc. Tutorials of the software can be found on the authors' website.

## 2.3 Benchmark and Exemplar Application

We evaluated *TaSer* and compared it with *varianttools* and *VariantAnnotation* on a desktop computer with 1 core of Xeon CPU X5660 2.80GHz. A dataset of variants on chromosome 1 from 1000 Genomes Project (1000 Genomes Project, et al., 2012) was used. The dataset consists of a total of ~3 million variants from 1,092 individuals (11 Gb after compression).

*TaSer* and *varianttools* follow a similar workflow, in that the whole dataset will first be processed in order to facilitate subsequent complex queries. Applying *TabAnno*, we annotated the VCF file using refFlat, a commonly used gene/transcript definition for UCSC genome browser, incorporating pre-calculated PolyPhen scores, and outputted a feature-enriched project file. The whole dataset was processed in 1.66 hours, and the peak memory usage is 43Mb. In order to use *varianttools*, a database project will need to be created. The process of building the database took 28.7 hours, and the peak memory usage is 648 Mb. It is clear that DBMS based method is computationally intensive in data preparation, which make it challenging to be applied on NGS datasets with many thousand samples.

We also evaluated both tools for making queries. Specifically, we extracted information of non-synonymous variants in 100 randomly selected genes in the feature-enriched VCF files, which took ~10 seconds for both tools.

We compared our tools with *VariantAnnotation*, an R-package that supports complex queries to VCF files, leveraging dynamically loaded bioinformatics databases. It does not require annotating the whole dataset to perform complex queries. Therefore it may be more flexible than our tools when it needs to switch to alternative annotations. However, when the same set of annotation is repeatedly used (e.g. in genetic association studies), *SeqMiner* can be more efficient for performing queries from feature-enriched files. For example, after loading necessary databases, *VariantAnnotation* took ~3.5 minutes to extract 100 genes on chromosome 1, which is >20 times slower than *SeqMiner* (10 seconds). Moreover, *SeqMiner* is more self-contained, requires fewer steps to perform the query, and stores results in standard R objects (e.g. matrix or list) for easier manipulation. An additional advantage of *TaSer* over *VariantAnnotation* is that it can calculate a variety of summary statistics for the entire NGS dataset besides making queries, such as transition/transversion ratio, which is an important metric for NGS data QC.

*TaSer* supports generic tab-delimited files, which represents a broader class of input formats than what *varianttools* and *VariantAnnotation* support. We also evaluated the performance of *TaSer* for annotating and querying generic tab-delimited files. Specifically, we simulated phenotype information, and generated METAL files of single site association test statistics for each variant in the 1000G dataset. The input files of summary statistics (251 Mb after bgzip compression) were annotated in 76 seconds and it took *SeqMiner* 7 seconds to extract summary statistics from 100 randomly chosen genes. It shows that our tools are capable of handling large datasets on a single desktop computer. *VariantAnnotation* and *varianttools* do not directly support generic tab-delimited files, and therefore were not compared on this dataset.

## 3 DISCUSSION AND CONCLUSION

We developed *TaSer*, a toolkit for annotating and querying NGS datasets. It complements existing software tools, and provides a valuable platform for statisticians to apply existing tools and develop novel methods for analyzing NGS data. Although it is mainly developed for analyzing human genetic data, it can also be used for other organisms, e.g. prokaryotes. If gene definitions are provided and properly formatted, our tools can be applied to annotate and query sequence datasets from these organisms as well. *TaSer* is currently being deployed to process NGS data from many thousands of individuals in our projects. We envision that it will continue to make valuable contributions to the analysis of NGS data.

## ACKNOWLEGEMENT

We would like to thank Drs. Gonçalo Abecasis, Hyun Min Kang Yanming Li for helpful discussions .

## REFERENCES

*1000 Genomes Project, C., et al. (2012) An integrated map of genetic variation from 1,092 human genomes, Nature,* **491***, 56-65.*

*Adzhubei, I.A., et al. (2010) A method and server for predicting damaging missense mutations, Nature methods,* **7***, 248-249.*

*Cooper, G.M., et al. (2005) Distribution and intensity of constraint in mammalian genomic sequence, Genome research,* **15***, 901-913.*

*Danecek, P., et al. (2011) The variant call format and VCFtools, Bioinformatics,* **27***, 2156-2158.*

*Kent, W.J., et al. (2002) The human genome browser at UCSC, Genome research,* **12***, 996-1006.*

*Li, H. (2011) Tabix: fast retrieval of sequence features from generic TAB-delimited files, Bioinformatics,* **27***, 718-719.*

*San Lucas, F.A., et al. (2012) Integrated annotation and analysis of genetic variants from next-generation sequencing studies with variant tools, Bioinformatics,* **28***, 421-422.*

*Willer, C.J., Li, Y. and Abecasis, G.R. (2010) METAL: fast and efficient meta-analysis of genomewide association scans, Bioinformatics,* **26***, 2190-2191.*